\documentclass[12pt]{article}
\begin{document}

\title{\bf Symmetries of the Energy-Momentum Tensor: Some Basic Facts}

\author{M. Sharif \thanks{e-mail: msharif@math.pu.edu.pk} and
Tariq Ismaeel \thanks{ismaeel$\_$tariq@yahoo.com}
\\ Department of Mathematics, University of the Punjab,\\ Quaid-e-Azam
Campus Lahore-54590, PAKISTAN.}

\date{}

\maketitle
\begin{abstract}
It has been pointed by Hall et al. [1] that matter collinations
can be defined by using three different methods. But there arises
the question of whether one studies matter collineations by using
the ${\cal L}_\xi T_{ab}=0$, or ${\cal L}_\xi T^{ab}=0$ or ${\cal
L}_\xi T_a^b=0$. These alternative conditions are, of course, not
generally equivalent. This problem has been explored by applying
these three definitions to general static spherically symmetric
spacetimes. We compare the results with each definition.
\end{abstract}


\section{Introduction}

It has been an interesting subject to use the symmetry group of a
spacetime in constructing the solution of Einstein field equation
(EFEs) given by
\begin{equation}
G_{ab}\equiv R_{ab}-\frac 12 Rg_{ab}=\kappa T_{ab},
\end{equation}
where $G_{ab}$ are the components of the Einstein tensor, $R_{ab}$
are the components of Ricci tensor and $T_{ab}$ are the components
of matter (energy-momentum) tensor, $R$ is the Ricci scalar and
$\kappa$ is the gravitational constant. Further, these solutions
are classified according to the Lie algebra structure generated by
these symmetries. The well-known connection between Killing
vectors (KVs) and constants of the motion [2,3] has encouraged the
search for general relations between collineations and
conservation laws [4]. Curvature and the Ricci tensors are the
important quantities which play a significant role in
understanding the geometric structure of spacetime. A pioneer
study of curvature collineations (CCs) and Ricci collineations
(RCs) has been carried out by Katzin, et al [5] and a further
classification of CCs and RCs has been obtained by different
authors [6,7].

The theoretical basis for the study of affine including Killing
and homothetic vector fields is well understood and many examples
are given. This is also true in the case of conformal fields as
well as projective and curvature collineations. However,
symmetries of the Ricci tensor and, in particular, energy-momentum
tensor have recently been studied. In this paper we shall analyze
the properties of a vector field along which the Lie derivative of
the energy-momentum tensor vanishes, i.e., ${\cal L}_\xi
T_{ab}=0$. But a natural question arises whether one studies
matter collineations defined in this way or those which satisfy
${\cal L}_\xi T^{ab}=0$, or ${\cal L}_\xi T_a^b=0$. These
different definitions are not generally equivalent. We shall apply
these three definitions to general static spherically symmetric
spacetimes to check this observation. Since the energy-momentum
tensor represents the matter part of the Einstein field equations
and gives the matter field symmetries. Thus the study of matter
collineations (MCs) seems more relevant from the physical point of
view.

There is a growing interest in the study of MCs [1,8-11 and
references therein]. Carot, et al [8] have discussed MCs from the
point of view of the Lie algebra of vector fields generating them
and, in particular, he discussed spacetimes with a degenerate
$T_{ab}$. Hall, et al [1], in the discussion of RC and MC, have
argued that the symmetries of the energy-momentum tensor may also
provide some extra understanding of the the subject which has not
been provided by KVs, Ricci and CCs. The same author also raised
the question how to define matter collineation. Keeping this point
in mind we address the problem of calculating MCs for static
spherically symmetric spacetimes using the three different
definitions. It is hoped that this would provide a better
understanding of MCs.

The distribution of the paper follows. In the next section, we
discuss some general issues about MCs and write down the MC
equations. In section three we calculate MCs by solving MC
equations for static spherically symmetric spacetimes using three
different conditions. Final section carries a discussion of the
results obtained.

\section{Some Basic Facts}

Let $(M,g)$ be a spacetime, $M$ being a Hausdorff, connected, four
dimensional manifold, and $g$ a Lorentz metric with signature
(+,-,-,-).

A vector $\xi$ is called a MC if the Lie derivative of the
energy-momentum tensor along that vector is zero. That is,
\begin{equation}
{\cal L}_\xi T=0,
\end{equation}
where $T$ is the energy-momentum tensor and ${\cal L_\xi}$ denotes
the Lie derivative along $\xi$ of the energy-momentum tensor $T$.
This equation, in a torsion-free space in a coordinate basis,
reduces to a partial differential equation,
\begin{equation}
T_{ab,c}\xi^c+T_{ac}\xi^c_{,b}+T_{bc}\xi^c_{,a}=0,\quad
a,b,c=0,1,2,3.
\end{equation}
where $,$ denotes partial derivative with respect to the
respective coordinate. We shall also consider those symmetries
generated by vector fields $\xi$ satisfying ${\cal L}_\xi
T^{ab}=0$ or ${\cal L}_\xi T_a^b=0$. These are ten coupled partial
differential equations for four unknown functions $(\xi^a)$ which
are functions of all spacetime coordinates in the case of
covariant and contravariant forms but sixteen for mixed form.

Collineations can be proper or improper. A collineation of a given
type is said to be {\it proper} if it does not belong to any of
the subtypes. When we solve MC equations, solutions representing
proper collineations can be found. However, in order to be related
to a particular conservation law, and its corresponding constants
of the motion, the {\it properness} of the collineation type must
be known.

We know that every KV is an MC, but the converse is not always
true. As given by Carot et al. [8], if $T_{ab}$ is non-degenerate,
$det(T_{ab})\neq 0$, the Lie algebra of the MCs is finite
dimensional. If $T_{ab}$ is degenerate, i.e., $det(T_{ab})=0$, we
cannot guarantee the finite dimensionality of the MCs. The study
of MCs has many difficulties which can be listed as follows
[1,12].

\begin{description}
\item{1.}  When we define affine and conformal vector fields on $M$, if the
vector field is at least $C^2$ and $C^3$ respectively, then $\xi$
is necessarily smooth on $M$. However, for any $k\in Z^+$ there
exists MC on spacetimes which are $C^k$ not $C^{k+1}$. The same is
true for the Ricci and CCs.

2.  An affine and a conformal vector field $\xi$ on $M$ is
uniquely determined by specifying it and its first covariant
derivative and specifying it and its first and second covariant
derivatives respectively at some $m\in M$. However, the value of
$\xi$ and its covariant derivatives of all orders at some $m\in M$
may not be enough to determine uniquely a MC $\xi$ on $M$. Thus
two MCs that agree on a non-empty open subset of $M$ may not agree
on $M$. These features are also found in RCs and CCS. This leads
to a problem of the extendibility of local MCs to the whole of $M$
which is more complicated than that for affine and conformal
vector fields [7].

3.  The set of all MCs on $M$ is a vector space but, like the set
of RCs or CCs and unlike the sets of affine and conformal vector
fields, it may be infinite dimensional and may not be a Lie
algebra. This latter defect arises from the fact that such
collineations must be $C^1$ in order that their definitions make
sense. But MCs (RCs and CCS) may be exactly $C^1$ and
differentiability may be destroyed under the Lie bracket
operation. If MCs are $C^\infty$ then one recovers the Lie algebra
structure but loses the non-smooth. The infinite dimensionality
may also lead to problems related to the orbits of the resulting
local diffeomorphism [7].

4.  If the energy-momentum tensor is of rank $4$, it may be
regarded as a metric on $M$. Then the family of $C^2$ MCs is, in
fact, a Lie algebra of smooth vector fields on $M$ of finite
dimension $\leq 10$ and $\neq 9$.
\end{description}
It is obvious from the EFEs (1) that left hand side is the
geometrical part constructed from the metric and its derivatives
while the right hand side is the physical part describing the
sources of the gravitational field. It is not clear whether (1) is
to be written  with the indices in the covariant, the
contravariant or the mixed positions in any case. These lead to
significant difficulties even with the definition of a matter
symmetry. It has been shown that for almost all spacetimes (in a
well defined topologically generic sense) the weyl tensor $C$ and
the energy-momentum tensor $T$ (or the Einstein tensor $G$)
determined the metric $g$ uniquely up to a constant conformal
factor and hence determined the Levi-Civita connection [13,14].
The special case of this result in vacuo is just Brinkmann's
theorem [13,15]. The following theorem can be considered as an
important result by considering the local diffeomorphisms
associated with a vector field $\xi$ on $M$ [1,13,14].\\
\par \noindent
{\bf Theorem}: Let $M$ be a spacetime manifold. Then, generically,
any vector field $\xi$ on $M$ which simultaneously satisfies
${\cal L}_\xi T=0~(\Leftrightarrow {\cal L}_\xi G=0)$ and ${\cal
L}_\xi C=0$ is a homothetic vector field.\\
Thus with this concept of symmetry for all the gravitational
sources, a metric symmetry (upto a constant homothetic scaling)
generically results.

The most general spherically symmetric metric is given as [15]
\begin{equation}
ds^2 = e^{\nu(t,r)}dt^2-e^{\lambda(t,r)}d
r^2-e^{\mu(t,r)}d\Omega^2,
\end{equation}
where $d\Omega^2=d\theta^2+\sin^2 \theta d\theta^2$. Since we are
dealing with static spherically symmetric spacetimes, Eq.(4)
reduces to
\begin{equation}
ds^2 = e^{\nu(r)}dt^2-e^{\lambda(r)}d r^2-e^{\mu(r)}d\Omega^2.
\end{equation}

\section{Solution of MC Equations Using Three Definitions}

In this section we shall use three different definitions to
calculate MCs of static spherically symmetric spacetimes.

\subsection{Solution When ${\cal L}_\xi T_{ab}=0$}

We can write MC Eqs.(3) in the expanded form as follows
\begin{equation}
T_{0,1} \xi^1 + 2 T_0 \xi^0_{,0} = 0,
\end{equation}
\begin{equation}
T_0 \xi^0_{,1} + T_1 \xi^1_{,0} = 0,
\end{equation}
\begin{equation}
T_0 \xi^0_{,2} + T_2 \xi^2_{,0} = 0,
\end{equation}
\begin{equation}
T_0 \xi^0_{,3} + \sin^2 \theta T_2 \xi^3_{,0} = 0,
\end{equation}
\begin{equation}
T_{1,1} \xi^1 + 2 T_1 \xi^1_{,1} = 0,
\end{equation}
\begin{equation}
T_1 \xi^1_{,2} + T_2 \xi^2_{,1} = 0,
\end{equation}
\begin{equation}
T_1 \xi^1_{,3} +\sin^2 \theta T_2 \xi^3_{,1} =0,
\end{equation}
\begin{equation}
T_{2,1} \xi^1 + 2 T_2 \xi^2_{,2} =0,
\end{equation}
\begin{equation}
\xi^2_{,3} + \sin^2 \theta \xi^3_{,2} =0,
\end{equation}
\begin{equation}
T_{2,1} \xi^1 + 2 T_2 \cot \theta \xi^2 + 2 T_2\xi^3_{,3} = 0,
\end{equation}
where $T_{3}=\sin^2\theta T_{2}$. It is to be noticed that we are
using the notation $T_{aa}=T_a$ etc. We solve these equations for
the degenerate as well as the non-degenerate case. The nature of
the solution of these equations changes when one (or more) $ T_{a}
$ is zero. The nature changes even if $ T_{a} \neq 0$ but $
T_{a,1} =0$.

A complete solution of MC Eqs.(6)-(15) has already been obtained
[11,17] using this covariant definition both for the degenerate as
well as non-degenerate cases. The degenerate case implies that
$det(T_{ab})=0$. When at least one of $T_a=0$, we can have the
following three main cases:\\
\par \noindent
\par \noindent
(1) when only one of the $T_a \neq 0$,
\par \noindent
\par \noindent
(2) when exactly two of the $T_a \neq 0$,
\par \noindent
\par \noindent
(3) when exactly three of the $T_a \neq 0$.\\
\par \noindent
\par \noindent
It is mentioned here that the trivial case, where $T_a=0$, shows
that every vector field is an MC. For the sake of comparison, we
would only give results in the form of tables at the end skipping
all the details as these can be seen elsewhere [11,17].

\subsection{Solution When ${\cal L}_\xi T^{ab}=0$}

In this case, MC equations can be written in the expanded form as
follows:
\begin{eqnarray}
T^0_{,1}\xi^1-2T^0\xi^0_{,0}=0, \\
T^0\xi^1_{,0}+T^1\xi^0_{,1}=0, \\
T^0\xi^2_{,0}+T^2\xi^0_{,2}=0, \\
T^0\sin^2\theta\xi^3_{,0}+T^2\xi^0_{,3}=0, \\
T^1_{,1}\xi^1-2T^1\xi^1_{,1}=0, \\
T^1\xi^2_{,1}+T^2\xi^1_{,2}=0, \\
T^1\sin^2\theta\xi^3_{,1}+T^2\xi^1_{,3}=0, \\
T^2_{,1}\xi^1-2T^2\xi^2_{,2}=0, \\
T^2(\sin^2\theta\xi^3_{,2}+\xi^2_{,3})=0, \\
\frac{T^2_{,1}}{2T^2}\xi^1-\cot\theta\xi^2-\xi^3_{,3}=0.
\end{eqnarray}
We solve this system of MC equations for the degenerate and
non-degenerate cases skipping the algebra.

\subsubsection{Degenerate Case}

The degenerate case implies the following three main cases:\\
\par \noindent
\par \noindent
(1) when only one of the $T^a \neq 0$,
\par \noindent
\par \noindent
(2) when exactly two of the $T^a \neq 0$,
\par \noindent
\par \noindent
(3) when exactly three of the $T^a \neq 0$.\\
\par \noindent
{\bf Case 1:} The first case gives either
$T^{0}\neq0,~T^{i}=0\quad(i=1,2,3)$ or
$T^{1}\neq0,~T^{j}=0\quad(j=0,2,3)$. When we use MC equations for
the case (1a), we further have two possibilities, i.e., either
$T^{0}=constant$ or $T^{0}\neq constant$. The first possibility
yields
\begin{equation}
\xi^{a}=\xi^{a}(r,\theta,\phi).
\end{equation}
The second option gives the following result
\begin{equation}
\xi^{0}=(\ln{\sqrt{T^{0}}})^{'}B(r,\theta,\phi)t+A(r,\theta,\phi),
\quad \xi^{i}=\xi^{i}(r,\theta,\phi).
\end{equation}
The case (1b), i.e., $T^{1}\neq0,\quad T^{j}=0$ gives the
following two options according to $T^1=constant$ or $T^1\neq
constant$. For $T^1=constant$, we obtain $\xi^{a}$ arbitrary. The
case $T^1\neq constant$ yields
\begin{equation}
\xi^{1}=\sqrt{T^{1}}B(t,\theta,\phi),\quad
\xi^{j}=\xi^{j}(t,\theta,\phi).
\end{equation}
\par \noindent
{\bf Case 2:} This case has the following two possibilities:\\
(2a) $T^{k}=0,\quad T^{l}\neq0$,\quad (2b) $T^{k}\neq0,\quad
T^{l}=0,\quad (k=0,1)(l=2,3)$ The first possibility gives either
$T^{2}=constant\neq0$ or $T^{2}\neq constant$. In the first case
we obtain
\begin{eqnarray}
\xi^{k}&=&C(t,r),\nonumber\\
\xi^{2}&=&A(t,r)\cos\phi+B(t,r)\sin\phi,\nonumber\\
\xi^{3}&=&\cot\theta[-A(t,r)\sin\phi+B(t,r)\cos\phi]+D(t,r).
\end{eqnarray}
In the second case, we have
\begin{eqnarray}
\xi^{0}&=&C(t,r)\nonumber,\\
\xi^{1}&=&0\nonumber,\\
\xi^{2}&=&A(t,r)\cos\phi+B(t,r)\sin\phi,\nonumber\\
\xi^{3}&=&\cot\theta[-A(t,r)\sin\phi+B(t,r)\cos\phi]+D(t,r).
\end{eqnarray}
The case (2b) yields the following options:\\
(i) $T^{0}=constant\neq0,\quad T^{1}=constant\neq0$,\\
(ii) $T^{0}=constant\neq0,\quad T^{1}\neq constant$,\\
(iii) $T^{0}\neq constant,\quad T^{1}=constant\neq0$,\\
(iv) $T^{0}\neq constant,\quad T^{1}\neq constant$.\\
For (2bi), we have
\begin{equation}
\xi^{a}=\xi^{a}(\theta,\phi).
\end{equation}
In the case (2bii), it follows that
\begin{equation}
\xi^{j}=\xi^{j}(\theta,\phi),\quad
\xi^{1}=\sqrt{T^{1}}A(\theta,\phi).
\end{equation}
The case (2biii) further yields three options according to the
value of $\alpha$ given by
\begin{equation}
\alpha=\frac{T^1}{T^0}(\frac{T^0_{,1}}{2T^0})'
\end{equation}
either $\alpha<0$, or $\alpha=0$, or $\alpha>0$. When $\alpha<0$,
we obtain the following solution
\begin{eqnarray}
\xi^{0}&=&\frac{(\ln{\sqrt{T^{0}}})'}{\sqrt{\alpha}}
[A(\theta,\phi)e^{\sqrt{\alpha}t}-B(\theta,\phi)e^{-\sqrt{\alpha}t}]
+C(\theta,\phi)\nonumber,\\
\xi^{1}&=&[A(\theta,\phi)e^{\sqrt{\alpha}t}
+B(\theta,\phi)e^{-\sqrt{\alpha}t}]\nonumber,\\
\xi^{l}&=&\xi^{l}(\theta,\phi).
\end{eqnarray}
For $\alpha=0$, the solution is given by
\begin{eqnarray}
\xi^{0}&=&(\ln{\sqrt{T^{0}}})'[A(\theta,\phi)\frac{t^{2}}{2}
+B(\theta,\phi)t]+C(\theta,\phi)\nonumber,\\
\xi^{1}&=&A(\theta,\phi)t+B(\theta,\phi)\nonumber,\\
\xi^{l}&=&\xi^{l}(\theta,\phi).
\end{eqnarray}
When $\alpha>0$, we have
\begin{eqnarray}
\xi^{0}&=&\frac{(\ln{\sqrt{T^{0}}})'}{\sqrt{\alpha}}
[A(\theta,\phi)\sin{\sqrt{\alpha}t}-B(\theta,\phi)
\cos{\sqrt{\alpha}t}]+C(\theta,\phi)\nonumber,\\
\xi^{1}&=&[A(\theta,\phi)\cos{\sqrt{\alpha}t}
+B(\theta,\phi)\sin{\sqrt{\alpha}t}]\nonumber,\\
\xi^{l}&=&\xi^{l}(\theta,\phi).
\end{eqnarray}
The last case, (2biv) when ${T^{0}}'\neq0,~{T^{1}}'\neq0$, further
yields three possibilities depending upon the value of $\beta$
given by
\begin{equation}
\beta=\frac{{\sqrt{T^{1}}[(\ln{\sqrt{T^{0}}})'\sqrt{T^{1}}}]'}{T^{0}},
\end{equation}
i.e., either $\beta<0$, or $\beta=0$, or $\beta>0$. The first
possibility implies that
\begin{eqnarray}
\xi^{0}&=&\frac{{(\ln{\sqrt{T^{0}}})'}\sqrt{T^{1}}}{\sqrt{\beta}}
[A_{1}(\theta,\phi)e^{\sqrt{\beta}t}-A_{2}
(\theta,\phi)e^{-\sqrt{\beta}t}]+A_{3}(\theta,\phi)\nonumber,\\
\xi^{1}&=&\sqrt{T^{1}}[A_{1}(\theta,\phi)e^{\sqrt{\beta}t}
+A_{2}(\theta,\phi)e^{-\sqrt{\beta}t}]\nonumber,\\
\xi^{l}&=&\xi^{l}(\theta,\phi).
\end{eqnarray}
The second option yields the following solution
\begin{eqnarray}
\xi^{0}&=&{(\ln{\sqrt{T^{0}}})'}{\sqrt{T^{1}}}
[A_{1}(\theta,\phi)\frac{t^{2}}{2}+A_{2}(\theta,\phi)t]
+A_{3}(\theta,\phi)\nonumber,\\
\xi^{1}&=&\sqrt{T^{1}}[A_{1}(\theta,\phi)t+A_{2}(\theta,\phi)]\nonumber,\\
\xi^{l}&=&\xi^{l}(\theta,\phi).
\end{eqnarray}
When $\beta>0$, we obtain
\begin{eqnarray}
\xi^{0}&=&\frac{{(\ln{\sqrt{T^{0}}})'}\sqrt{T^{1}}}{\sqrt{\beta}}
[A_{1}(\theta,\phi)\sin{\sqrt{\beta}t}-A_{2}(\theta,\phi)
\cos{\sqrt{\beta}t}]+A_{3}(\theta,\phi)\nonumber,\\
\xi^{1}&=&\sqrt{T^{1}}[A_{1}(\theta,\phi)\cos{\sqrt{\beta}t}
+A_{2}(\theta,\phi)\sin{\sqrt{\beta}t}]\nonumber,\\
\xi^{l}&=&\xi^{l}(\theta,\phi).
\end{eqnarray}
{\bf Case 3:} In this case we have three of $T^{a}\neq0$ which
further yields two cases.
\begin{eqnarray*}
(3a)\quad T^{0}=0,~T^{i}\neq0,\quad (3b)\quad T^{1}=0,~T^{j}\neq0.
\end{eqnarray*}
When we take the first case (3a), MC equations give the following
four possibilities:
\begin{eqnarray*}
(3ai)\quad T^{1}=constant,~T^{2}=constant,\\
(3aii)\quad T^{1}\neq constant,~T^{2}=constant,\\
(3aiii)\quad T^{1}=constant,~T^{2}\neq constant,\\
(3aiv)\quad T^{1}\neq constant,~T^{2}\neq constant.
\end{eqnarray*}
For (3ai), we obtain the following solution
\begin{eqnarray}
\xi^{k}&=&B(t),\nonumber\\
\xi^{2}&=&[A_{1}(t)\cos{\phi}+A_{2}(t)\sin{\phi}]\nonumber,\\
\xi^{3}&=&-\cot\theta[A_{1}(t)\sin{\phi}-A_{2}(t)\cos{\phi}]+A_{3}(t).
\end{eqnarray}
The case (3aii) gives the following result
\begin{eqnarray}
\xi^{0}&=&A_{1}(t),\quad
\xi^{1}=\sqrt{T^{1}}A_{2}(t),\nonumber\\
\xi^{2}&=&A_{3}(t)\cos{\phi}+A_{4}(t)\sin{\phi},\nonumber\\
\xi^{3}&=&-\cot\theta[A_{3}(t)\sin{\phi}-A_{4}(t)\cos{\phi}]+A_{5}(t).
\end{eqnarray}
The cases (3aiii) and (3aiv) yield the same solution given by
\begin{eqnarray}
\xi^{0}&=&A_1(t),\quad \xi^{1}=0,\nonumber\\
\xi^{2}&=&A_2(t)\cos{\phi}+A_3(t)\sin{\phi}\nonumber,\\
\xi^{3}&=&-\cot\theta[A_2(t)\sin{\phi}-A_3(t)\cos{\phi}]+A_4(t).
\end{eqnarray}
The case (3b) also yields four possibilities given by
\begin{eqnarray*}
(3bi)\quad T^{0}=constant,~T^{2}=constant,\\
(3bii)\quad T^{0}\neq constant,~T^{2}=constant,\\
(3biii)\quad T^{0}=constant,~T^{2}\neq constant,\\
(3biv)\quad T^{0}\neq constant,~T^{2}\neq constant.
\end{eqnarray*}
The case (3bi) gives the following solution
\begin{eqnarray}
\xi^{0}&=&A_{1}(r),\quad
\xi^{1}=A_{2}(r)\nonumber,\\
\xi^{2}&=&[A_{3}(r)\cos{\phi}+A_{4}(r)\sin{\phi}]\nonumber,\\
\xi^{3}&=&-\cot\theta[A_{3}(r)\sin{\phi}-A_{4}(r)\cos{\phi}]+A_5(r).
\end{eqnarray}
For the case(3bii), it implies that
\begin{eqnarray}
\xi^{0}&=&(\ln{\sqrt{T^{0}}})'A_{1}(r){t}+A_{2}(r),\quad
\xi^{1}=A_{1}(r),\nonumber\\
\xi^{2}&=&[A_{3}(r)\cos{\phi}+A_{4}(r)\sin{\phi}],\nonumber\\
\xi^{3}&=&-\cot\theta[A_{3}(r)\sin{\phi}-A_{4}(r)\cos{\phi}]+A_{5}(r).
\end{eqnarray}
The third and fourth possibilities give the same result
\begin{eqnarray}
\xi^{0}&=&A_{1}(r),\quad \xi^{1}=0,\nonumber\\
\xi^{2}&=&[A_{2}(r)\cos{\phi}+A_{3}(r)\sin{\phi}],\nonumber\\
\xi^{3}&=&-\cot\theta[A_{2}(r)\sin{\phi}-A_{3}(r)\cos{\phi}]+A_4(r).
\end{eqnarray}

\subsubsection{Non-Degenrate Case}

Solving Eqs.(16)-(20), we have
\begin{equation}
\xi^1=\sqrt{T^1}A(t,\theta,\phi)
\end{equation}
with
\begin{equation}
\ddot{A}(t,\theta,\phi)=-\alpha A(t,\theta,\phi),\quad
\alpha=\frac{\sqrt{T^1}}{T^0} (\frac{{T^0}'\sqrt{T^1}}{2{T^0}})'.
\end{equation}
There arises three possibilities:
\begin{equation}
1.\quad \alpha<0, \quad 2. \quad \alpha=0,\quad 3.\quad \alpha>0.
\end{equation}
{\bf Case 1:} In this case we take $\alpha=-\alpha$ so that we
have
\begin{equation}
A(t,\theta,\phi)=A_1(\theta,\phi)\cosh\sqrt{\alpha}t
+A_2(\theta,\phi)\sinh\sqrt{\alpha}t.
\end{equation}
Using this value of $A$ in the remaining MC equations, we get the
following two cases, i.e. either $\frac{T^2}{T^0}=constant$ or
$\frac{T^2}{T^0}\neq constant$.

In the first case (1a), MCs are given by
\begin{eqnarray}
\xi^0&=&c_{0},\quad \xi^1=0,\quad
\xi^2=c_{1}\cos\phi+c_{2}\sin\phi \nonumber,\\
\xi^3&=&\cot\theta(-c_{1}\sin\phi+c_{2}\cos\phi)+c_{3}.
\end{eqnarray}
Using the case (1b) together with MC equations we obtain the
following two cases
\begin{eqnarray*}
(1bi)\quad A_{1,2}(\theta,\phi)=0=A_{2,2}(\theta,\phi),\quad\quad
\quad\quad\quad\quad\quad\quad\quad\quad\nonumber\\
(1bii)\quad(\frac{-2\alpha B
+\beta}{2\alpha})(\frac{T^2}{T^1})'+\frac{T^2}{\sqrt{T^1}}=0,\quad
B=\int{\frac{T^0}{\sqrt{T^1}}dr},
\end{eqnarray*}
where $\beta$ is an integration constant. For (1bi), we have the
following solution
\begin{eqnarray}
\xi^0&=&\frac{{T^0}'\sqrt{T^1}}{2T^0{\sqrt{\alpha}}}
(c_{4}\sinh{\sqrt{\alpha}{t}}+c_{5}\cosh{\sqrt{\alpha }{t}})
+ c_{0}\nonumber,\\
\xi^1&=&\sqrt{T^1}(c_{4}\cosh{\sqrt{\alpha}{t}}
+c_{5}\sinh{\sqrt{\alpha}{t}})\nonumber,\\
\xi^2&=&c_{1}\cos\phi+c_{2}\sin\phi \nonumber,\\
\xi^3&=&\cot\theta(-c_{1}\sin\phi+c_{2}\cos\phi)+c_{3}
\end{eqnarray}
with $T^2=constant$. For $T^2\neq constant$, it reduces to (1a).
The case (1bii) gives
\begin{equation}
\eta=\frac{{T^2}'\sqrt{T^1}\alpha}{T^2\epsilon(-2\alpha
B+\beta)^2},
\end{equation}
where
\begin{eqnarray*}
\epsilon=\frac{T^2}{T^0(-2\alpha B+\beta)}.
\end{eqnarray*}
Eq.(53) implies that either (1bii*) $\eta>0$, where $\eta$ can be
1 or not 1, or (1bii**) $\eta<0$. In the case (1bii*+) MCs are
\begin{eqnarray}
\xi^0&=&\frac{(-2\alpha B+\beta)}{2\sqrt{\alpha}}
[\{c_{4}\cos{\theta}+(c_{6}\cos\phi
+c_{7}\sin{\phi})\sin{\theta}\}\sinh{\sqrt{\alpha}{t}}\nonumber\\
&+&\{c_{5}\cos{\theta}+(c_{8}\cos{\phi}
+c_{9}\sin{\phi})\sin{\theta}\}\cosh{\sqrt{\alpha}{t}}]+ c_{0},\nonumber\\
\xi^1&=&\sqrt{T^1}[\{c_{4}\cos{\theta}+(c_{6}\cos\phi
+c_{7}\sin{\phi})\sin{\theta}\}\cosh{\sqrt{\alpha}{t}}\nonumber\\
&+&\{c_5\cos{\theta}+(c_{8}\cos{\phi}+c_{9}\sin{\phi})
\sin{\theta}\}\sinh{\sqrt{\alpha}{t}}],\nonumber\\
\xi^2&=&-\frac{\epsilon(-2\alpha B+\beta)^{2}}{2{\alpha}}
[\{-c_{4}\sin{\theta}+(c_{6}\cos\phi+c_{7}\sin{\phi})
\cos{\theta}\}\cosh{\sqrt{\alpha}{t}}\nonumber\\
&+&\{-c_{5}\sin{\theta}+(c_{8}\cos{\phi}
+c_{9}\sin{\phi})\cos{\theta}\}\sinh{\sqrt{\alpha
}{t}}]+c_{1}\cos\phi+c_{2}\sin\phi, \nonumber\\
\xi^3&=&\frac{-\epsilon(-2\alpha
B+\beta)^{2}\csc\theta}{2{\alpha}}[(-c_{6}\sin\phi
+c_{7}\cos{\phi})\cosh{\sqrt{\alpha}{t}}\nonumber\\&+&(-c_{8}\sin{\phi}
+c_{9}\cos{\phi})\sinh{\sqrt{\alpha}{t}}]
+\cot\theta(-c_{1}\sin\phi+c_{2}\cos\phi)+c_{3}.
\end{eqnarray}
The cases (1bii*++) and (1bii**) reduce to (1a).\\
{\bf Case 2:} This case gives
$\frac{{T^0}'\sqrt{T^1}}{2{T^0}}=constant$ which implies that
either (2a) $constant=0$ or (2b) $constant\neq0$. The case (2a)
yields $T^0=constant$ which together with MC equations implies
that either (2ai) $T^2=constant$ or (2aii) $T^2\neq constant$. It
follows from (2ai) that either $T^1=constant$ or $T^1\neq
constant$. The case (2ai*) gives the following MCs
\begin{eqnarray}
\xi^0=-a c_{4}\frac{r}{\sqrt{b}}+c_{0},\quad\xi^1=\sqrt{b}(c_{4}t
+c_{5}),\nonumber\\
\xi^2=c_{1}\cos\phi+c_{2}\sin\phi, \quad
\xi^3=\cot\theta(-c_{1}\sin\phi+c_{2}\cos\phi)+c_{3}.
\end{eqnarray}
For (2ai**), we obtain the following solution
\begin{eqnarray}
\xi^0=-a c_{4}\int{\frac{dr}{\sqrt{T^{1}}}}+c_{0},\quad
\xi^1=\sqrt{T^{1}}(c_{4}t+c_{5}),\nonumber\\
\xi^2=c_{1}\cos\phi+c_{2}\sin\phi, \quad
\xi^3=\cot\theta(-c_{1}\sin\phi+c_{2}\cos\phi)+c_{3}.
\end{eqnarray}
The case (2aii) yields the the same solution as for the case (1a).

The case (2b) further gives the following two possibilities
\begin{eqnarray}
(T^2)'-\frac{2cT^2}{\sqrt{T^1}}=0,\quad(T^2)'-\frac{2cT^2}{\sqrt{T^1}}\neq0
\end{eqnarray}
The case (2bi) reduces to (1a). For (2bii), we obtain
\begin{eqnarray}
\xi^0=\epsilon_{1}(c_{4}\frac{t^2}{2}+c_{5}t)
-\frac{c_{4}\lambda_{1}e^{2\epsilon_{1}\int{\frac{dr}
{\sqrt{T^{1}}}}}}{2\epsilon_{1}}+c_{0},\quad
\xi^1=\sqrt{T^{1}}(c_{4}t+c_{5}),\nonumber\\
\xi^2=c_{1}\cos\phi+c_{2}\sin\phi, \quad
\xi^3=\cot\theta(-c_{1}\sin\phi+c_{2}\cos\phi)+c_{3},
\end{eqnarray}
where $\epsilon_{1}=\frac{{T^0}'\sqrt{T^1}}{2{T^0}}=constant$.\\
{\bf Case 3:} This case is very similar to the case 1 and can be
solved on the same lines.

\subsection{Solution When ${\cal L}_\xi T^a_b=0$}

For this definition, MC equations take the following expanded form
\begin{eqnarray}
{{T^{0}}_{0}}_{,1}\xi^{1}=0,\\
({{T^{0}}_{0}}-{{T^{1}}_{1}}){\xi^{0}}_{,1}=0,\\
({{T^{0}}_{0}}-{{T^{2}}_{2}}){\xi^{0}}_{,2}=0,\\
({{T^{0}}_{0}}-{{T^{3}}_{3}}){\xi^{0}}_{,3}=0,\\
({{T^{1}}_{1}}-{{T^{0}}_{0}}){\xi^{1}}_{,0}=0,\\
{{T^{1}}_{1}}_{,1}\xi^{1}=0,\\
({{T^{1}}_{1}}-{{T^{2}}_{2}}){\xi^{1}}_{,2}=0,\\
({{T^{1}}_{1}}-{{T^{3}}_{3}}){\xi^{1}}_{,3}=0,\\
({{T^{2}}_{2}}-{{T^{0}}_{0}}){\xi^{2}}_{,0}=0,\\
({{T^{2}}_{2}}-{{T^{1}}_{1}}){\xi^{2}}_{,1}=0,\\
{{T^{2}}_{2}}_{,1}\xi^{1}=0,\\
({{T^{3}}_{3}}-{{T^{0}}_{0}}){\xi^{3}}_{,0}=0,\\
({{T^{3}}_{3}}-{{T^{1}}_{1}}){\xi^{3}}_{,1}=0,\\
{{T^{3}}_{3}}_{,1}\xi^{1}=0.
\end{eqnarray}
When we solve these MC equations, after some algebra, we obtain
arbitrary MCs for all the possibilities of degenerate and
non-degenerate cases [18].

\section{Conclusion}

We know that the metric tensor is non-degenerate whereas the
Ricci, Riemann and energy-momentum tensors are not necessarily
non-degenerate. When there is a degeneracy, it is possible to have
arbitrary collineations. Thus KVs will always be definite but
collineations can be indefinite. Further, if the relevant tensor
vanishes, all vectors become collineations as for Minkowski space
where every vector is a MC. Also, for vacuum spacetime every
vector will be an MC as it is Ricci flat (e.g. Schwarzschild
metric).

There is a problem with respect to the definition of MC because of
the possible choices $T_{ab},~T^{ab},~T^a_b$. In this paper we
have evaluated MCs for static spherically symmetric spacetimes
using the three definitions. The motivation behind this is two
fold: First to check whether these three definitions give similar
results. If not the same which one gives the more interesting
results. This would help us to understand the distribution of
matter symmetries and comparison to the symmetries of the metric,
Ricci and curvature tensors. We discuss the results
obtained using the table given below:\\
\par \noindent
\par \noindent
{\bf {\small Table 1.}} {\bf MCs using ${\cal L}_\xi
T_{ab}=0$,~${\cal L}_\xi T^{ab}=0$ or ${\cal L}_\xi T_a^b=0$ for
the Degenerate Case}

\vspace{0.1cm}

\begin{center}
\begin{tabular}{|l|l|l|l|}
\hline {\bf Cases} & {\bf ${\cal L}_\xi T_{ab}=0$} & {\bf ${\cal
L}_\xi T^{ab}=0$} & {\bf ${\cal L}_\xi T_a^b=0$}
\\ \hline 1ai & Infinite-dimensionl& Infinite-dimensionl& Infinite-dimensionl
\\ \hline 1aii & Infinite-dimensionl&
Infinite-dimensionl & Infinite-dimensionl
\\ \hline 1b & Infinite-dimensionl &
Infinite-dimensionl & Infinite-dimensionl
\\ \hline 2ai & Infinite-dimensionl &
Infinite-dimensionl & Infinite-dimensionl
\\ \hline 2aii & Infinite-dimensionl &
Infinite-dimensionl & Infinite-dimensionl
\\ \hline 2bi & Infinite-dimensionl&
Infinite-dimensionl & Infinite-dimensionl
\\ \hline 2bii & Infinite-dimensionl & Infinite-dimensionl & Infinite-dimensionl
\\ \hline 2biii & Infinite-dimensionl & Infinite-dimensionl & Infinite-dimensionl
\\ \hline 3ai & Infinite-dimensionl & Infinite-dimensionl & Infinite-dimensionl
\\ \hline 3aii$*$ & Infinite-dimensionl & Infinite-dimensionl & Infinite-dimensionl
\\ \hline 3aii$**$ & Infinite-dimensionl & Infinite-dimensionl & Infinite-dimensionl
\\ \hline 3aiii & Infinite-dimensionl & Infinite-dimensionl & Infinite-dimensionl
\\ \hline 3bi & $4$ & Infinite-dimensionl & Infinite-dimensionl
\\ \hline 3bii & Infinite-dimensionl & Infinite-dimensionl & Infinite-dimensionl
\\ \hline 3biii & Infinite-dimensionl & Infinite-dimensionl & Infinite-dimensionl
\\ \hline 3biv & Infinite-dimensionl & Infinite-dimensionl & Infinite-dimensionl
\\ \hline
\end{tabular}
\end{center}
\par \noindent
\par \noindent
{\bf {\small Table 2.}} {\bf MCs using ${\cal L}_\xi
T_{ab}=0$,~${\cal L}_\xi T^{ab}=0$ or ${\cal L}_\xi T_a^b=0$ for
the Non-degenerate Case}

\vspace{0.1cm}

\begin{center}
\begin{tabular}{|l|l|l|l|l|}
\hline {\bf Cases} & {\bf ${\cal L}_\xi T_{ab}=0$} & {\bf Cases} &
{\bf ${\cal L}_\xi T^{ab}=0$} & {\bf ${\cal L}_\xi T_a^b=0$}
\\ \hline 1a & $6$ & 1a & $4$ & Infinite-dimensionl
\\ \hline 1bi & $4$ & 1bi &
$6$ & Infinite-dimensionl
\\ \hline 1bii$*$ & $6$ & 1bii$*+$ &
$10$ & Infinite-dimensionl
\\ \hline 1bii$**$ & $6$ & 1bii$*++$ &
$4$ & Infinite-dimensionl
\\ \hline 1biii & $4$ & 1bii$**$ &
$4$ & Infinite-dimensionl
\\ \hline 2ai & $4$ & 2ai$*$ &
$6$ & Infinite-dimensionl
\\ \hline 2aii & $4$ & 2ai$**$ & $6$ & Infinite-dimensionl
\\ \hline 2b & $10$ & 2aii, 2bi & $4$ & Infinite-dimensionl
\\ \hline 2c & $4$& 2bii & $6$ & Infinite-dimensionl
\\ \hline 3 & $4$ & 3 & Same as 1 & Infinite-dimensionl
\\ \hline
\end{tabular}
\end{center}
From these tables we see that out of three definitions, the mixed
form gives an arbitrary MC in all cases. The contravariant form
yields infinite dimensional MCs for the degenerate case but finite
for the non-degenerate case. For the covariant definition, there
exists an interesting case which gives finite number of MCs in one
possibility even for the degenerate case. The non-degenerate case
gives finite number of MCs in covariant form. We know that if
$T_{ab}$ is non-degenerate, the Lie algebra of the MCs is finite
dimensional but if $T_{ab}$ is degenerate, finite dimensionality of
MCs  cannot be guaranteed. The covariant and contravariant
definitions verify this statement but the mixed form yields infinite
dimensional MCs even for the non-degenerate case. It is always an
interesting feature if there exist finite MCs in the degenerate case
and the covariant definition gives such a possibility. This feature
obviously motivates one to use this definition for the
classification of spacetimes according to MCs. It is worth
mentioning here that the purely covariant and contravariant cases
give exactly the same class of MCs for the non-degenerate case.

We would like to comment the degenerate case a little more. It has
been shown that, in general, there are infinity many MCs which musty
be found by the solution of relevant MC equations. However, the MCs
in the degenerate case are not as useful as the ones of the
non-degenerate case. Indeed the assumption of the degeneracy of
$T_{ab}$ leads to differential equations which fix the metric
functions up to arbitrary constant of integration. Hence the form of
the matter tensor can be determined making the constraint imposed
the MC redundant. It has been shown [9] that only interesting case
for the degenerate case is when rank $T_{ab}=1$, i.e., a null
Einstein-Maxwell field or a dust fluid.

One may ask whether there can exist vectors which are simultaneously
solutions of ${\cal L}_\xi T_{ab}=0$ and ${\cal L}_\xi T^{ab}=0$ or
of ${\cal L}_\xi T^b_a=0$. We see from table $1$, for the degenerate
case, the only common solutions of ${\cal L}_\xi T_{ab}=0$ and
${\cal L}_\xi T^{ab}=0$ are the arbitrary collineations and for the
non-degenerate case (table $2$) the only common solutions are
necessarily isometries. For ${\cal L}_\xi T_{ab}=0$ and ${\cal
L}_\xi T^b_a=0$, the only common solution are the arbitrary
collineations (degenerate case) but there is no common solution for
the non degenerate case.

The infinite dimensionality of MCs may lead to different problems,
e.g., one cannot define Lie algebra. Also, this may lead to problems
related to the orbits of the resulting local diffeomorphism [7, 12].
In the light of such problems and the usefulness of the covariant
definition for MC [10,11] it can be concluded that the covariant
definition of MC should be
preferred.\\

{\bf \large Acknowledgment}\\

One of the authors (MS) would like to thank Prof. Graham S. Hall for
the useful discussion on the topic.

\vspace{0.5cm}

{\bf \large References}

\begin{description}

\item{[1]} Hall, G.S., Roy, I. and Vaz, L.R.: Gen. Rel and Grav. {\bf 28}(1996)299.

\item{[2]} Noether, E.: Nachr, Akad. Wiss. Gottingen, II, Math. Phys. {\bf K12}(1918)235.

\item{[3]} Davis, W.R. and Katzin, G.H.: Am. J. Phys. {\bf 30}(1962)750.

\item{[4]} Bokhari, A.H. and Qadir, Asghar: J. Math. Phys. {\bf
34}(1993)3543; J. Math. Phys. {\bf 31}(1990)1463.

\item{[5]} Katzin, G.H., Levine, J. and Davis, H.R.: J. Math. Phys.{\bf 10}(1969)617.

\item{[6]} Bokhari, A.H., Kashif, A.R. and Qadir, Asghar: J. Math. Phys. {\bf
41}(2000)2167.

\item{[7]} Hall, G.S. and da Costa, J.: J. Math. Phys. {\bf 32}(1991)2848;2854.

\item{[8]} Carot, J. and da Costa, J.: {\it Procs. of the 6th Canadian Conf. on
General Relativity and Relativistic Astrophysics}, Fields Inst.
Commun. 15, Amer. Math. Soc. WC Providence, RI(1997)179.

\item{[9]} Carot, J., da Costa, J. and Vaz, E.G.L.R.: J. Math. Phys. {\bf 35}(1994)4832.

\item{[10]} Camci, U. and Sharif, M.: Gen. Rel. and Grav. {\bf
35}(2003)97;\\
Camci, U. and Sharif, M.: Class. Quantum Grav. {\bf
20}(2003)2169;\\
Sharif, M.:  J. Math. Phys. {\bf 44}(2003)5141; J. Math. Phys.
{\bf 45}(2004)1518; J. Math. Phys. {\bf 45}(2004)1532.

\item{[11]} Sharif, M. and Aziz, Sehar: Gen. Rel. and Grav. {\bf
35}(2003)1093.

\item{[12]} Hall, G.S.: {\it Symmetries and Curvature Structure in General Relativity}
(World Scientific, 2004).

\item{[13]} Hall, G.S.: {\it Lecture at Conference on Differential Geometry}
(University of Coimbra, Portugal, 1984).

\item{[14]} Hall, G.S. and Rendall, A.D.: J. Math. Phys. {\bf 28}(1987)1837.

\item{[15]} Ehlers, J. and Kundt, W.: {\it In Gravitation: An
Introduction to Current Research,} ed. Witten, L. (Wiley, New
York, 1962).

\item{[16]} Stephani, H., Kramer, D., Maccallum, M., Hoenselaers, C. and Herlt,E.
            \textit{Exact Solutions of Einstein's Field Equations}
            (Cambridge University Press, 2003).

\item{[17]} Aziz, Sehar: M.Phil. Thesis (University of the Punjab,
2002).

\item{[18]} Ismaeel, Tariq: M.Phil. Thesis (University of the Punjab,
2006).

\end{description}

\end{document}